\documentclass[fleqn,10pt,onecolumn]{article}
\usepackage[utf8]{inputenc}
\usepackage[T1]{fontenc}
\usepackage{graphics}
\usepackage{bm}
\usepackage{authblk}
\usepackage{amsmath}
\usepackage{graphicx}
\usepackage[margin=1in]{geometry}

\title{Viscoelastic coarsening of quasi-2D foam} 

\author[1]{Chiara Guidolin}
\author[2]{Jonatan Mac Intyre}
\author[1]{Emmanuelle Rio}
\author[2]{Antti Puisto}
\author[1]{Anniina Salonen}
\affil[1]{Université Paris-Saclay, CNRS, Laboratoire de Physique des Solides, Orsay 91405, France}
\affil[2]{Aalto University, Department of Applied Physics, Espoo 02150, Finland}

\affil[*]{e-mail: chiara.guidolin@universite-paris-saclay.fr,  anniina.salonen@universite-paris-saclay.fr}

\begin{document}

\flushbottom
\maketitle

\begin{abstract}
\textbf{Foams are unstable jammed materials. They evolve over timescales comparable to their "time of use", which makes the study of their destabilisation mechanisms crucial for applications. In practice, many foams are made from viscoelastic fluids, which are observed to prolong their lifetimes. Despite their importance we lack understanding of the coarsening mechanism in such systems. We probe the effect of continuous phase viscoelasticity on foam coarsening with foamed emulsions. We show that bubble size evolution is strongly slowed down and foam structure hugely impacted. The main mechanisms responsible are the absence of continuous phase redistribution and a non-trivial link between foam structure and mechanical properties. These combine to give spatially heterogeneous coarsening. Beyond their importance in the design of foamy materials, the results give a macroscopic vision of phase separation in a viscoelastic medium.}

\end{abstract}

Foams coarsen as gas diffuses between bubbles due to differences in Laplace pressure.
The rate of coarsening is set by foam topology through the organisation of the neighbouring bubbles\cite{vonNeumann_Coarsen,Kraynik2001}.
The growth law is theoretically predicted and experimentally verified in two limiting cases of very dry or very wet foams.
In both cases the foam reaches a self-similar scaling state with invariant statistical distributions \cite{Mullins1986_2D}, which leads to a power-law growth of the mean bubble size $R$.
In dry foams gas diffusion occurs exclusively through the thin films separating the bubbles and $R(t) \propto t^{1/2}$,
while in dilute bubble dispersions the gas transfer through the bulk phase is captured within a mean-field description and  $R(t) \propto t^{1/3}$\cite{Lifshitz1961}. 

If gas diffusion is no longer the rate-limiting step in the coarsening process, the evolution could be different. In wet foams made from viscoelastic fluids, lamellar\cite{Zenaida2015} or smectic\cite{Buchanan2002}, exponents below $1/3$ have been measured. 

Bubbles in a viscoelastic continuous phase stop coarsening if the surrounding material is sufficiently stiff to oppose the growth (or shrinking) of the bubbles.
The conditions for arrest have been predicted for both single bubbles and foams \cite{Kloek2001_Bubble, Webster2001_Coarsen}.
Bey \textit{et al.} worked with monodisperse bubbles to show that, indeed, if the elastic energy of the medium is higher than the surface energy coarsening is halted \cite{Bey2017_StopCoarsen}.
Lesov \textit{et al.} measured an arrest once the continuous phase yield stress was higher than Laplace pressure \cite{Lesov2014}.

The relevant parameter to compare surface tension effects to bulk elasticity is the elastocapillary number $Ca_\text{el}$
\cite{Kogan2013}.
In the case of capillary inclusions (liquid drops or gas bubbles) in an elastic matrix, $Ca_\text{el}$ is defined as the ratio of the bulk elastic modulus $G_0$ to Laplace pressure, thus $Ca_\text{el} = \frac{G_0}{\gamma/R}$, where $\gamma$ is the surface tension and $R$ the bubble (drop) size.
Depending on $Ca_\text{el}$, the inclusions can either soften or stiffen the material \cite{Ducloue2014_Inclusions,Style2015_ElastoInclusions}.
In a foam the bubbles cannot be considered as spherical inclusions, as they are deformed with thin films between them.
However $Ca_\text{el}$ has been shown to be the relevant parameter to describe foam elasticity due to the additivity of the bubble and continuous phase contributions \cite{Gorlier2017_ElastoFoams,Mikhailovskaya2020_ElastoCapFoam}.

In practice, we encounter foamy materials in a broad range of $Ca_\text{el}$. Between the limits of capillary-controlled foams (e.g. soap froths) and solid foams, lies a number of materials with intermediate $Ca_\text{el}$. These are foams made from soft solids (creams, pastes, gels) or precursors of solid foams. In these systems, $Ca_\text{el}$ evolves in time, due to an increase of the bubble size (coarsening, coalescence) or solidification of the continuous phase, both of which lead to an increase of $Ca_\text{el}$.
Understanding bubble size evolution in such systems is crucial as it is a key control parameter of the material properties \cite{Gibson1997}.
We thus explore the evolution of the bubble size and foam structure in a foam made from a model viscoelastic fluid (an emulsion), in which we vary the initial elastocapillary number from 1 to 8. In this range we expect the continuous phase to impact the coarsening process without arresting it.

\begin{figure}[htb]
    \centering
    \includegraphics[width=0.5\columnwidth]{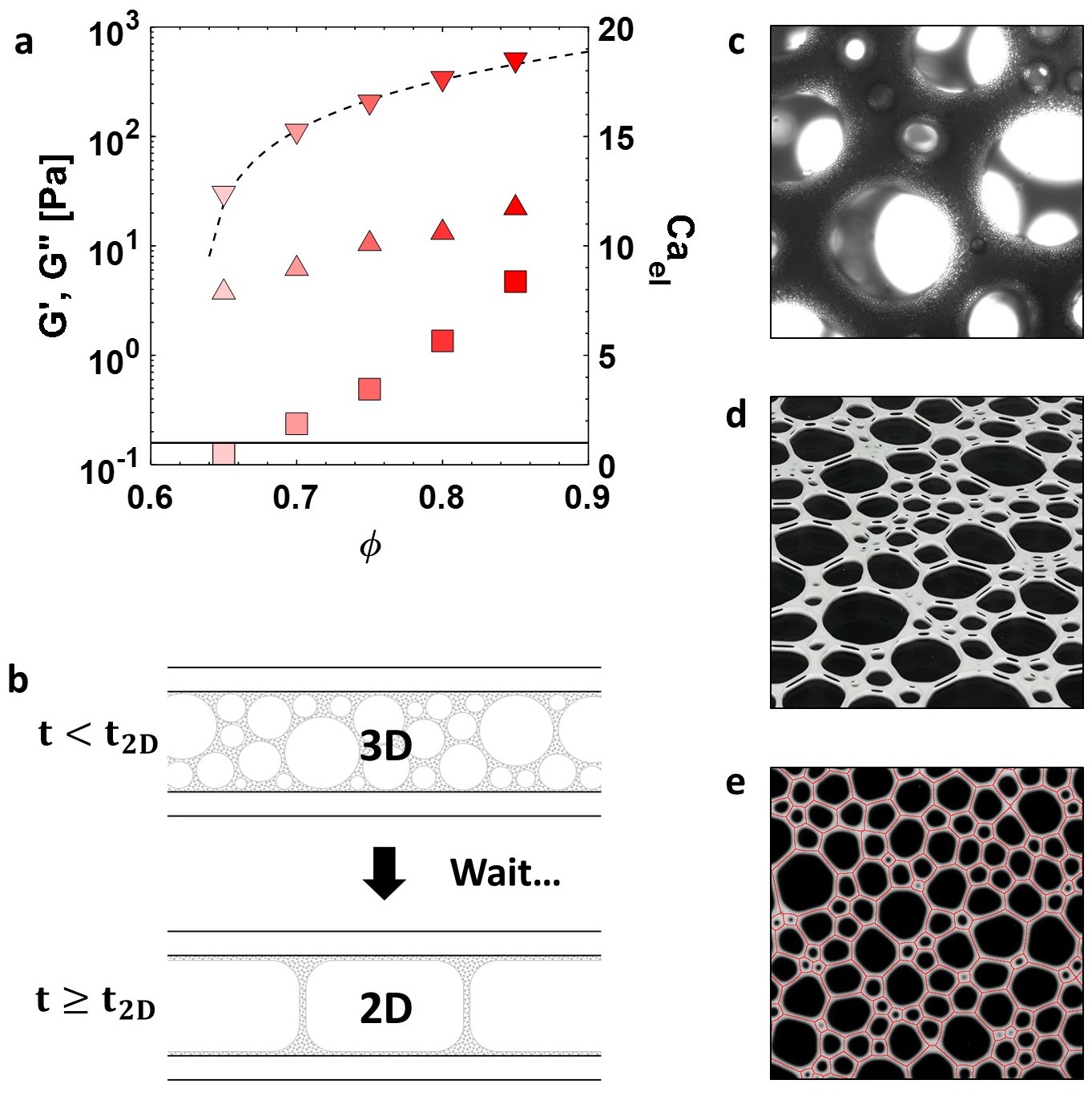}
    \caption{\textbf{Foamed emulsion setup. a}, Storage (down triangles) and loss moduli (up triangles) of the emulsions used as the foam continuous phases as a function of oil volume fraction $\phi$. The dashed line shows the expected scaling of $G' \approx \phi(\phi - \phi^*)$ with $\phi^*=0.635$ from \cite{Mason1995}, which describes the data well. In the following we use $G_0$ for $G'$ for the emulsion elastic modulus. The squares are $Ca_\text{el}$ calculated using $\langle R \rangle $ = 0.5 mm and a surface tension of 30 mN/m and the solid line guides to $Ca_\text{el} = 1$.  
    \textbf{b}, Schematic drawing of the sample generation from 3D foam to the quasi-2D foam. The gap between the plates is 1 mm. 
    \textbf{c}, Photograph of a foam after generation seen under the microscope. The edge size is 830 $\mu$m.
    \textbf{d}, Perspective view of a quasi-2D foam.
    \textbf{e}, Top view of a foam with its skeleton overlaid in red. The edge size is 30 mm.}
    \label{fig:ExpApproach}
\end{figure}

In concentrated emulsions the elastic and viscous moduli can be easily varied through changing the volume fraction $\phi$ of the dispersed phase\cite{Mason1995}. 
This is why we use oil-in-water emulsions, with an average drop size of 2 $\mu$m and $\phi$ between 0.65 to 0.85, which allows us to vary the continuous phase elastic modulus from 31 to 506 Pa (Fig. \ref{fig:ExpApproach}a).
All the studied emulsions are predominantly elastic, the loss modulus always being less than 10\% of the storage modulus.
We use sodium dodecyl sulphate (SDS) at 30 g/L as the surfactant to stabilise both the oil/water and the gas/water interfaces.
The choice of such a high SDS concentration ensures that all the interfaces are well covered. 

We choose to work with quasi-2D foams, which consist of a single layer of bubbles sandwiched between two glass plates (Fig. \ref{fig:ExpApproach}d).
Such systems have been widely used by the foam community to explore many aspects of foam stability\cite{schimming2017,Forel2019} and rheology\cite{kabla2003}.
The coarsening laws are the same as in 3D foams, but the specific configuration allows the tracking of individual bubbles for a fine analysis of the foam structure\cite{Gay2011_2DStructure}.
Instead of creating the foam directly between the plates, we first make a 3D foamed emulsion with small polydisperse bubbles and a liquid (emulsion) fraction $\epsilon \approx 11 \%$ using a mixer to incorporate air in the emulsion (Fig. \ref{fig:ExpApproach}c).
Then we fill and close the cell to leave the bubbles to grow. 
This allows us to start with quasi-2D foams with small, polydisperse bubbles (Fig. \ref{fig:ExpApproach}d).
The images are skeletonised to measure the bubble size (Fig. \ref{fig:ExpApproach}e).
We use $\langle R\rangle$ = 0.5 mm to calculate the initial $Ca_\text{el}$, which for the different samples increases from 0.5 to 8 with increasing $\phi$ (Fig. \ref{fig:ExpApproach}a).

\begin{figure}[htb]
    \centering
    \includegraphics[width=0.5\columnwidth]{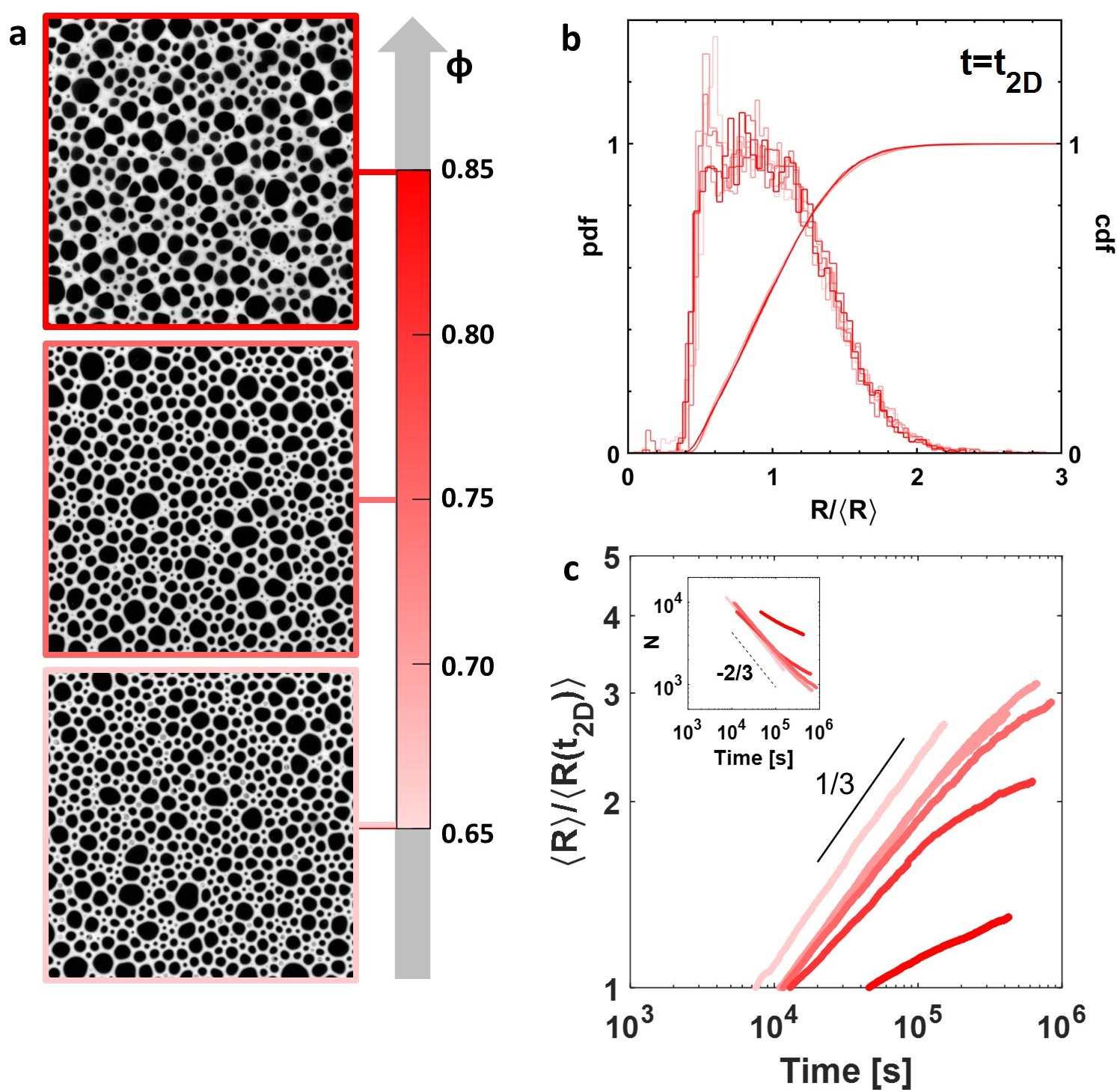}
    \caption{\textbf{Initial foams and size evolution. a}, Photographs of foams made from the different emulsions at $t=t_\text{2D}$. From top to bottom: $\phi$=0.85, 0.75, 0.65. The edge of each frame is 40 mm.
    \textbf{b}, Bubble size distributions at $t=t_\text{2D}$ for the five different foams studied, on the left axis the probability distribution function and on the right axis the cumulative distribution function. The initial distributions of the foams are almost identical, and the cumulative distributions overlap almost perfectly.
    \textbf{c}, Normalised mean bubble size growth. The black solid line is a $1/3$ powerlaw, which is the lower limit for diffusion limited coarsening. Inset: Temporal evolution of the total number of measured bubbles, where the limit for diffusive ripening is given by the $-2/3$ law.}
    \label{fig:Figure2}
\end{figure}

Photographs of three foams at different oil fractions, taken at $t = t_\text{2D}$, are shown in Fig. \ref{fig:Figure2}a.
The pictures at $\phi = 0.65$ and $0.75$ are similar and the typical cellular foam structure can be recognised.
But at $\phi = 0.85$ the foam already has some peculiar features, such as elongated bubble shapes, which we will return to later on.
Despite the morphological differences, the bubble size distributions at this point are very similar in all the foams considered.
This can be seen from the distribution functions of the dimensionless radius $R/\langle R\rangle$ (Fig. \ref{fig:Figure2}b), which are also comparable to the steady state distributions measured in aqueous quasi-2D foams \cite{Glazier1990}.

The temporal evolution of the average radius normalised by the average radius at $t = t_\text{2D}$ of the quasi-2D foams changes considerably with $\phi$ (Fig. \ref{fig:Figure2}c).
The foam made from the least elastic emulsion at $\phi = 0.65$ evolves the most rapidly and with an apparent power law. However, the exponent is smaller than the expected $t^{1/2}$ for dry foams and closer to $t^{1/3}$ expected for very wet foams, although the foam is rather dry.
This is a consequence of finite size effects observed in moderately dry quasi-2D foams, where the vertical film size decreases as the foam coarsens at constant liquid fraction \cite{Glazier}.
Even though our systems are dry enough for adjacent bubbles to share thin films, all samples have coarsening rates equal to or below the prediction for dilute bubble dispersions. This is the lowest possible rate for diffusion-limited growth, and any change from diminished gas transport

in the emulsion would only shift the curves, but not change the power laws observed. 
The foams made from emulsions at higher $\phi$ evolve very slowly, and with continuously slowing coarsening rates with increasing bubble size.
As the liquid fraction in the foams is very similar, the differences between the samples are due to the emulsion. 

\begin{figure*}[htb]
    \centering
    \includegraphics[width=\textwidth]{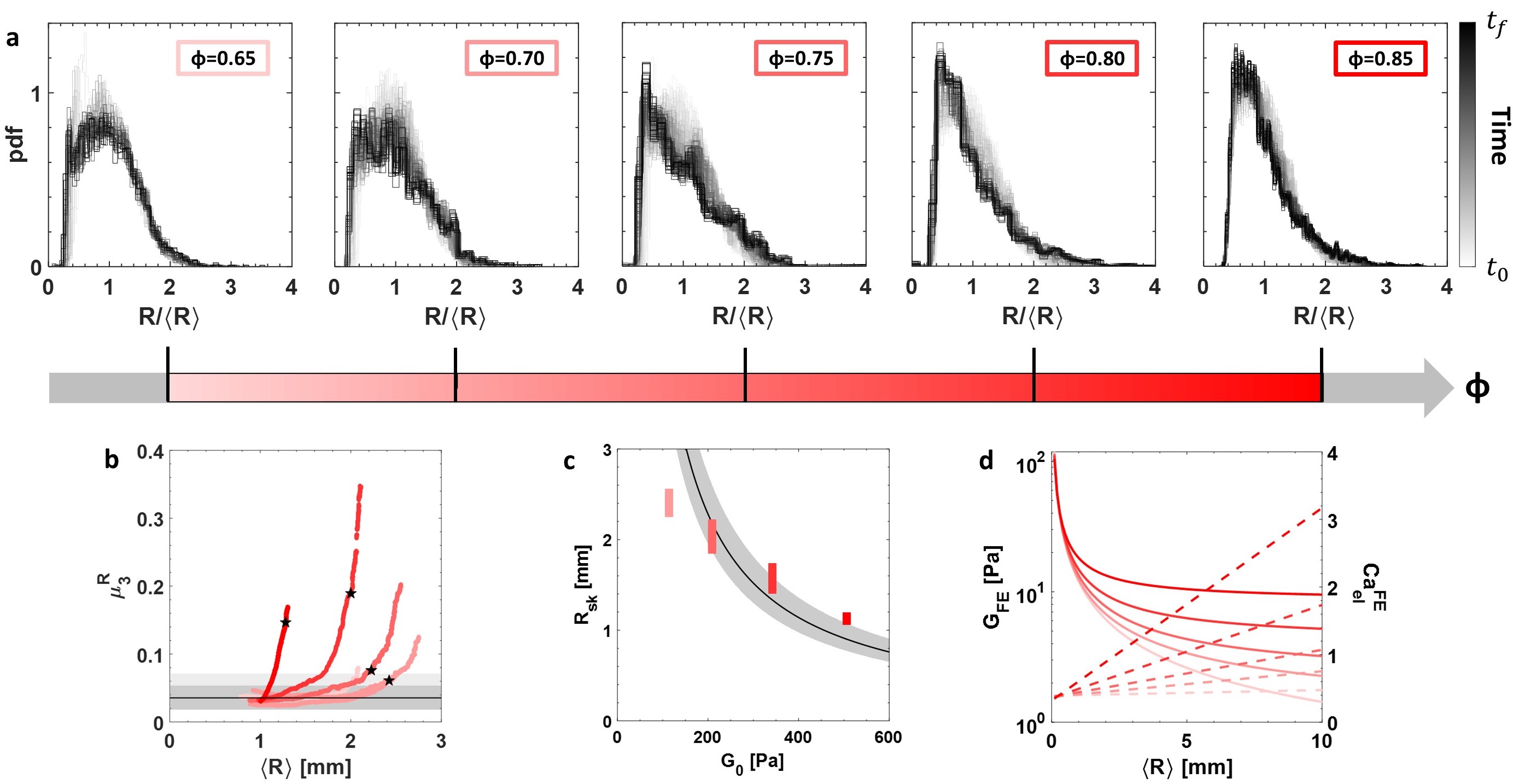}
    \caption{ \textbf{Evolution of bubble size distributions. a}, Time evolution of the dimensionless bubble size distributions in the foams made from emulsions at different $\phi$. The curves are plotted with a grey scale proportional to the foam age, where black corresponds to the end of the image acquisition $t_f$.
    \textbf{b}, Third moment $\mu_3^R$ versus the mean bubble radius. The solid black line indicates the average value at early times. The grey zones show the $\pm 50\% $ and $+100\%$ values, which are used to estimate $R_\text{sk}$ in d. The $\mu_3^R$ have black stars, which indicate the values at $t = 96$ hours.
    \textbf{c}, Impact of emulsion elastic modulus on $R_\text{sk}$. The black line shows the prediction with $Ca_\text{el}^\text{FE} = 0.54$ and the grey region the impact of changing liquid fraction on it.
    \textbf{d}, Evolution of the expected foam elastic modulus $G_\text{FE}$ (solid lines) and of the foam elastocapillary number $Ca_\text{el}^\text{FE}$ (dashed lines) as the mean bubble size grows because of coarsening, for the different $\phi$.}
    \label{fig:Figure3}
\end{figure*}

We have a high number of bubbles $N$ throughout the coarsening process (inset of Fig. \ref{fig:Figure2}c), which allows us to measure bubble size distributions in time with good statistics. 
We thus compare the evolution of the dimensionless bubble size distributions in the foams made from the different emulsions (Fig. \ref{fig:Figure3}a).
It is clear that the elasticity of the continuous phase has an impact on the distributions, which unlike aqueous foams, do not head towards self-similar distributions obtained at steady state\cite{Glazier1990}.
Although the exact evolution depends on $\phi$, in all of them a shift of the peak to smaller $R/\langle R \rangle$ can be observed
This shift implies an accumulation of smaller bubbles along time, which indicates a delay in their disappearance that is at the origin of the slow coarsening evolutions observed.

We choose to quantify the change in the shape of the distributions through its third central moment $\mu_3^R$, which is sensitive to the distribution asymmetry or skew and, in this case, to the presence of small bubbles.
By plotting $\mu_3^R$ as a function of the average bubble radius (Fig. \ref{fig:Figure3}b), we can see that $\mu_3^R$ is initially around 0.03 for all the foams,
but as the bubble size increases it departs from this plateau.
The higher $\phi$, the smaller the bubble size at which the skew deviates.
We note that at $\phi$ = 0.65 a deviation is observed as the bubbles reach around 2 mm, but this is due to the onset of coalescence and hence the physical origin is different than in the other samples.
We define the radius at which $\mu_3^R$ has increased by 50\% from its average value (from 0.036 to 0.053) as $R_\text{sk}$.
This gives us a characteristic radius at which the foam evolution has deviated from that of classical foams. $R_\text{sk}$ decreases with increased $\phi$, and hence higher $G_0$ (Fig. \ref{fig:Figure3}c).

We expect the departure from standard foam coarsening to happen once the elasticity of the continuous phase becomes large compared to Laplace pressure.
However, in a foam the bubbles are not surrounded by a continuous emulsion phase of constant elastic modulus, but by other bubbles forming a foam. The elastic modulus of a 3D foam with an elastic continuous phase can be described by an additive model combining bubble $G_F$ and continuous phase $G_E$ contributions as $G_\text{FE} = G_F + G_E$\cite{Gorlier2017_ElastoFoams}. The foam elasticity is modeled by $G_F = 1.6 \dfrac{\gamma}{\langle R\rangle}(1-\epsilon)(0.36 - \epsilon) \equiv f(\epsilon) \frac{\gamma}{\langle R\rangle}$ \cite{Cohen-Addad2013_FlowinFoams}, and the continuous phase by $G_E = \epsilon^2 G_0$ \cite{Gibson1997}, where $\epsilon$ is the foam liquid fraction. Despite the structural differences between quasi-2D and 3D foams we use the 3D foam description to capture the dependency of $G_{FE}$ on $R$ and $G_0.$ 
For simplicity we do not use the coupling function introduced by Gorlier et al\cite{Gorlier2017_ElastoFoams}, as it only has a weak effect on our estimated moduli and as the function could be system dependent.
The elastic moduli of foams $G_\text{FE}$ calculated with the foam liquid fraction $\epsilon$ and the elastic moduli $G_0$ corresponding to our emulsions, are shown as a function of the average bubble radius in Fig. \ref{fig:Figure3}d.
At small $R$ the $G_\text{FE}$ are all close as they are dominated by the elasticity of the bubbles, while at larger bubble sizes the foam elasticity depends strongly on $\phi$. 

As the foams coarsen the average bubble size increases and elastic effects from the emulsion become relatively more important.
In order to compare the latter to Laplace pressure, we can construct a second elastocapillary number as $Ca_\text{el}^\text{FE} = G_\text{FE}/(\gamma/R) = f(\epsilon) + \langle R\rangle G_0 \epsilon ^2 / \gamma$, which increases with $\langle R\rangle$ (Fig. \ref{fig:Figure3}d). It has been shown that monodisperse foams stop coarsening once $Ca_\text{el}^\text{FE} \approx 0.8$ \cite{Bey2017_StopCoarsen}. We can describe the evolution of $R_\text{sk}$ with $G_0$ assuming a critical $Ca_\text{el}^\text{FE} = 0.54$ (Fig. \ref{fig:Figure3}c) unique to all $\phi$. Therefore, in our foams elastic effects from the continuous phase become visible in the bubble size distributions once $2G_\text{FE} \approx \gamma / \langle R\rangle$. The experiment at $\phi = $ 0.65 stops at $Ca_\text{el}^\text{FE} \approx 0.4$, below the critical $Ca_\text{el}^\text{FE}$, which is why for these foams the distribution are unchanged (Fig. \ref{fig:Figure3}a). The foams also never stop coarsening despite reaching $Ca_\text{el}^\text{FE} = 0.8$, however this is because they are polydisperse.

\begin{figure*}[h]
    \centering
    \includegraphics[width=\textwidth]{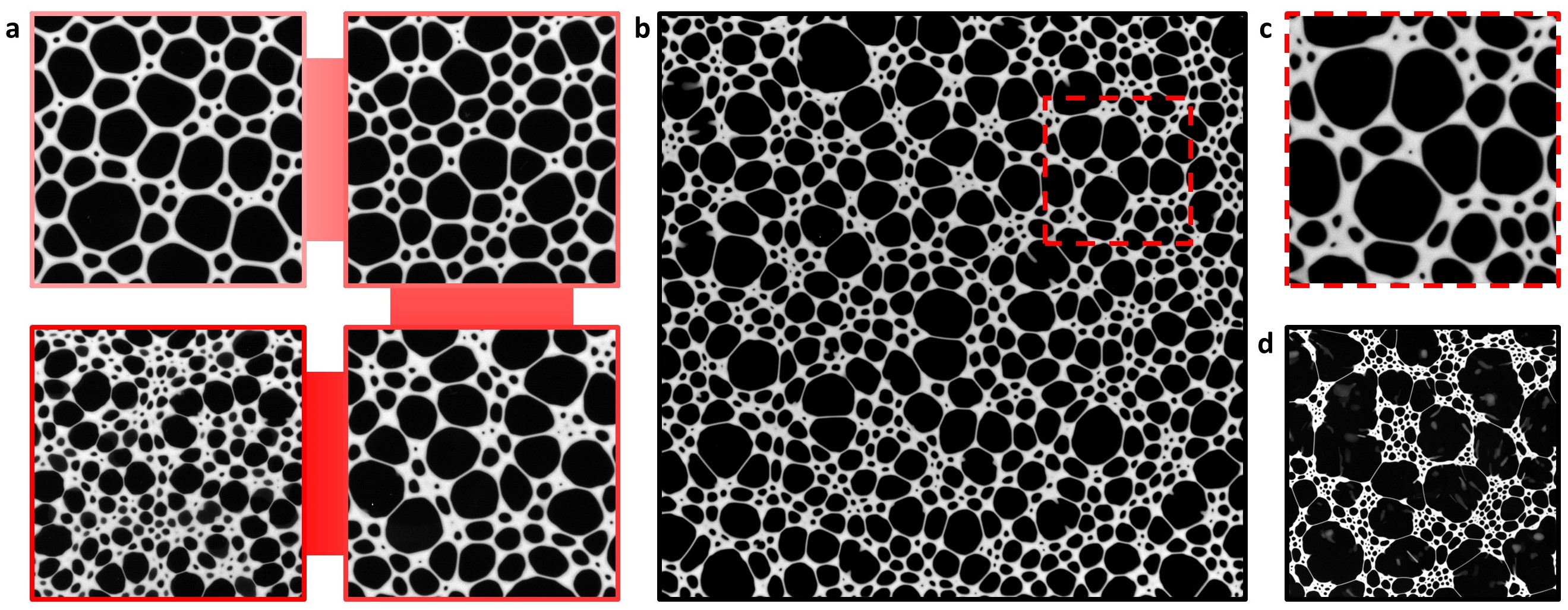}
    \caption{\textbf{Foamed emulsion structure.} \textbf{a}, Sample structure after 96 hours at different $\phi$. From top left to bottom left, clockwise: $\phi$=0.70, 0.75, 0.80, 0.85. The edge of each frame is 4 cm. The corresponding skew of the distributions is indicated by a star in Fig. \ref{fig:Figure3}b. At higher $\phi$ the foam structure becomes highly unconventional with unrelaxed bubble shapes and uneven emulsion distribution. 
    \textbf{b}, Sample at $\phi$=0.80 after 5 days. Chains of large bubbles and regions of small bubbles become clearly visible. Edge size 12 cm.
    \textbf{c}, Enlargement showing the emulsion bulge at the ends of the thin Plateau borders between large bubbles. Edge size 3 cm.
    \textbf{d}, Sample having $\phi$=0.80 made with sunflower oil, after 12 days. The edge of the frame is 13 cm. Segregation into regions of large and small bubbles is complete.  }
    \label{fig:Figure4}
\end{figure*}

Bubble size polydispersity in our samples also plays a crucial role in their structural evolution. 
Since $G_\text{FE}$ decreases with $R$ (Fig. \ref{fig:Figure3}d), in a polydisperse foam larger bubbles are easier to deform than smaller ones.
This spatial heterogeneity in the foam mechanical properties affects the coarsening process causing a peculiar evolution of the foam structure. Bubbles will grow preferentially towards weaker regions, thus towards other large bubbles. The resulting heterogeneous ageing, which leads to a segregation between smaller and larger bubbles, becomes visible after 96 hours of ageing (Fig. \ref{fig:Figure4}a) for samples with $\phi \geq 0.75$. The clusters of small bubbles enclosed by the chains of larger ones are striking at $\phi$ = 0.8 (Fig. \ref{fig:Figure4}b.)

The emulsion elasticity also changes the structure more locally.
We observe the emulsion bulging into the bubbles at the vertices (Fig. \ref{fig:Figure4}c). Such inversion of curvature is another sign of
the easier deformability of the bubbles instead of the emulsion, and departure from capillary mediated Plateau's laws.

The inhomogeneous evolution of the foam due to the heterogeneity in mechanical properties with bubble size resembles the elastic ripening recently observed by Rosowski \textit{et al.} at a microscopic scale \cite{Rosowski2020_NatPhys}.
In their work ripening is arrested in elastic regions, and continues in the less elastic ones, where the inhomogeneity is created by changing the polymer matrix while in our case the inhomogeneity is inherently due to the variation in the response of the foam with bubble size. 

If the foams are left to evolve for a very long time, the larger bubbles will keep growing until they start to coalesce, leaving the islands of small bubbles behind (Fig. \ref{fig:Figure4}d).
The photograph is reminiscent of structures observed during the final stages of viscoelastic phase separation identified by Tanaka\cite{Tanaka2000} in systems with high dynamic asymmetry.
He proposed its occurrence in foams despite the difference in the coarsening processes from the polymer phase separation or colloidal gelation previously studied, and the images could be a first indication of such a process at millimetric length scales.

We show that viscoelasticity of the foam continuous phase leads to a slowing down of coarsening and a radical change in foam structure. Once a critical $Ca_\text{el}^\text{FE}$ is reached, the bubble size distributions show the accumulation of smaller bubbles, which eventually segregate from the larger ones. This is caused by the interplay between bubble size and foam elasticity, which combined with bubble size polydispersity results in highly heterogeneous materials. Stress build up in the unrelaxed structures will be transferred to the foams if dried or solidified resulting in mechanical weakness. The deviation from the traditional cellular foam structure can have other implications on the material properties, so understanding and control of the structural evolution is crucial for the safe design of foamy materials.

\section*{MATERIALS AND METHODS}

\subsection*{Materials}
Air bubbles and oil droplets (rapeseed oil from Brassica Rapa, Sigma-Aldrich) are dispersed in an aqueous solution containing 30 g/L of sodium dodecyl sulfate (SDS, from Sigma-Aldrich).

\subsection*{Emulsion generation}
Concentrated O/W emulsions are generated by mechanically mixing the oil and the aqueous surfactant solution with the double syringe method.
This technique allows to control the oil volume fraction and ensures a good reproducibility of the samples.
The resulting droplet size distribution is measured with laser diffraction with a Mastersizer 3000E (Malvern Panalytical) equipped with a Hydro SM wet dispersion unit.
The size distribution is polydisperse, with a span, defined as $d(90\%)-d(10\%)/d(50\%)$, ranging from 1.2 to 1.
The surface-weighted mean drop diameter ranges between 3 and 5 $\mu$m.
Both the average size and the distribution width decrease with increasing $\phi$, consistently with emulsification in turbulent flow \cite{Tcholakova2011a}.

\subsection*{Emulsion rheology}
The mechanical properties of concentrated emulsions are measured with a MCR302 rheometer (Anton Paar).
The emulsion storage and loss moduli are obtained by performing oscillatory strain sweep tests (frequency 1 Hz, amplitude from 10$^{-5}$ to 1) in a cylindrical Couette geometry (CC27), with a gap of 1.1 mm.
All measurements are performed at (20.3 $\pm$ 0.1)$^{\circ}$C.

\subsection*{Foam generation}
Emulsions are foamed with a planetary kitchen mixer (Kenwood MultiOne 1000 W).
The mixing speed is gradually increased from the minimum to the maximum level, and after 25 minutes of whipping, up to roughly 90\% of air is incorporated inside the emulsion, without affecting the droplet size distribution.
The final foam liquid fraction is measured by weight: a glass container of known volume $V_\text{foam}$ is filled with freshly-made foamed emulsion and then weighed, so that, if the gas density is assumed to be zero, the mass of the foam liquid content $m_\text{em}$ is measured.
We assume the emulsion density to be given by the weighted sum of the density of its two components, namely $\rho_\text{em} \simeq \phi \rho_\text{oil} + (1-\phi)\rho_\text{water}$, so that the foam liquid fraction is retrieved through the relation $\varepsilon = m_\text{em}/ (\rho_\text{em} V_\text{foam})$.
The final foam liquid fraction ranges between 9\% and 13\%, and slightly increases with $\phi$, consistently with the increase of emulsion bulk viscosity \cite{Politova2018}.

\subsection*{Imaging and image treatment}
Freshly-generated foamed emulsions are carefully sandwiched between two square glass plates (edge length 24 cm), separated by rubber joint of thickness 1 mm which sets the cell gap.
The cell is then placed between two square metal frames which are screwed to keep the cell closed.
The cell is then put under a digital camera (Basler acA3800 - 14um, resolution 3840x2748 pixel) equipped with a lens (TAMRON 16mm f/1.4), and a square of LED lights provides rather uniform illumination from above.
A photo of the foamed emulsion is taken every 3 minutes at early stage, and then every 30 minutes at late coarsening stage.

Custom-made MATLAB scripts are used for processing the images.
A first image pre-treatment is carried out by cropping the raw frames around a region of interest, adjusting the contrast and obtaining the 2D foam skeleton through a watershed algorithm.
The skeletonised frames are then processed with a second MATLAB script based on the built-in function \textit{regionprops} to retrieve the area of each gas cell, from which the equivalent radius is calculated as $R=\sqrt{A/\pi}$.
\\

\bibliography{references} 

\begin{thebibliography}{10}

\bibitem{vonNeumann_Coarsen}
J.~Von~Neumann.
\newblock {\em Metal Interfaces (American Society for Metals, Cleveland,
  1952),}.

\bibitem{Kraynik2001}
Andrew~M. Kraynik, Stephan~A. Koehler, Howard~A. Stone, and Sascha Hilgenfeldt.
\newblock {An accurate von Neumann's law for three-dimensional foams}.
\newblock {\em Physical Review Letters}, 86(12):2685--2688, 2001.

\bibitem{Mullins1986_2D}
W.~W. Mullins.
\newblock {The statistical self-similarity hypothesis in grain growth and
  particle coarsening}.
\newblock {\em Journal of Applied Physics}, 59(4):1341--1349, 1986.

\bibitem{Lifshitz1961}
I.M. Lifshitz and V.V. Slyozov.
\newblock The kinetics of precipitation from supersaturated solid solutions.
\newblock {\em Journal of Physics and Chemistry of Solids}, 19(1):35--50, 1961.

\bibitem{Zenaida2015}
Zenaida Brice{\~{n}}o-Ahumada, Amir Maldonado, Marianne Imp{\'{e}}ror-Clerc,
  and Dominique Langevin.
\newblock {On the stability of foams made with surfactant bilayer phases.}
\newblock {\em Soft matter}, pages 1459--1467, 2016.

\bibitem{Buchanan2002}
Mark Buchanan.
\newblock {Liquid Crystal Foams: Formation and Coarsening}.
\newblock {\em arXiv}, page 0206477, 2002.

\bibitem{Kloek2001_Bubble}
W~Kloek, T~van Vliet, and M~Meinders.
\newblock {Effect of bulk and interfacial rheological properties on bubble
  dissolution}.
\newblock {\em Journal of Colloid and Interface Science}, 237(2):158--166,
  2001.

\bibitem{Webster2001_Coarsen}
A~J Webster and M~E Cates.
\newblock {Osmotic Stabilization of Concentrated Emulsions and Foams}.
\newblock {\em Langmuir}, 62(17):595--608, 2001.

\bibitem{Bey2017_StopCoarsen}
Houda Bey, Fr{\'{e}}d{\'{e}}ric Wintzenrieth, Olivier Ronsin, Reinhard
  H{\"{o}}hler, and Sylvie Cohen-Addad.
\newblock {Stabilization of foams by the combined effects of an insoluble gas
  species and gelation}.
\newblock {\em Soft Matter}, 13(38):6816--6830, 2017.

\bibitem{Lesov2014}
I.~Lesov, S.~Tcholakova, and N.~Denkov.
\newblock {Factors controlling the formation and stability of foams used as
  precursors of porous materials}.
\newblock {\em Journal of Colloid and Interface Science}, 426:9--21, 2014.

\bibitem{Kogan2013}
M~Kogan, L~Ducloue, J~Goyon, X~Chateau, O~Pitois, and G~Ovarlez.
\newblock {Mixtures of foam and paste: suspensions of bubbles in yield stress
  fluids}.
\newblock {\em Rheologica Acta}, 52(3):237--253, 2013.

\bibitem{Ducloue2014_Inclusions}
Lucie Duclou{\'{e}}, Olivier Pitois, Julie Goyon, Xavier Chateau, and Guillaume
  Ovarlez.
\newblock {Coupling of elasticity to capillarity in soft aerated materials}.
\newblock {\em Soft Matter}, 10(28):5093, 2014.

\bibitem{Style2015_ElastoInclusions}
Robert~W. Style, Rostislav Boltyanskiy, Benjamin Allen, Katharine~E. Jensen,
  Henry~P. Foote, John~S. Wettlaufer, and Eric~R. Dufresne.
\newblock {Stiffening solids with liquid inclusions}.
\newblock {\em Nature Physics}, 11(1):82--87, 2015.

\bibitem{Gorlier2017_ElastoFoams}
F~Gorlier, Y~Khidas, and O~Pitois.
\newblock {Journal of Colloid and Interface Science Coupled elasticity in soft
  solid foams}.
\newblock {\em Journal of Colloid and Interface Science}, 501:103--111, 2017.

\bibitem{Mikhailovskaya2020_ElastoCapFoam}
Alesya Mikhailovskaya, V{\'{e}}ronique Trappe, and Anniina Salonen.
\newblock {Colloidal gelation, a means to study elasto-capillarity effects in
  foam}.
\newblock {\em Soft Matter}, pages 2249--2255, 2020.

\bibitem{Gibson1997}
L.J. Gibson and M.F. Ashby.
\newblock {\em {Cellular solids: structure and properties}}.
\newblock Cambridge edition, 1997.

\bibitem{Mason1995}
T.~G. Mason, J.~Bibette, and D.~A. Weitz.
\newblock {Elasticity of compressed emulsions}.
\newblock {\em Physical Review Letters}, 1995.

\bibitem{schimming2017}
C.~D. Schimming and D.~J. Durian.
\newblock Border-crossing model for the diffusive coarsening of two-dimensional
  and quasi-two-dimensional wet foams.
\newblock {\em Phys. Rev. E}, 96:032805, Sep 2017.

\bibitem{Forel2019}
Emilie Forel, Benjamin Dollet, Dominique Langevin, and Emmanuelle Rio.
\newblock Coalescence in two-dimensional foams: A purely statistical process
  dependent on film area.
\newblock {\em Phys. Rev. Lett.}, 122:088002, Feb 2019.

\bibitem{kabla2003}
Alexandre Kabla and Georges Debr{\'e}geas.
\newblock Local stress relaxation and shear banding in a dry foam under shear.
\newblock {\em Physical review letters}, 90(25):258303, 2003.

\bibitem{Gay2011_2DStructure}
C~Gay, P~Rognon, D~Reinelt, and F~Molino.
\newblock {Rapid Plateau border size variations expected in three simple
  experiments on 2D liquid foams}.
\newblock {\em European Physical Journal E}, 34(1), 2011.

\bibitem{Glazier1990}
James~A. Glazier, Michael~P. Anderson, and Gary~S. Grest.
\newblock {Coarsening in the two-dimensional soap froth and the large- Q potts
  model: A detailed comparison}.
\newblock {\em Philosophical Magazine B: Physics of Condensed Matter;
  Statistical Mechanics, Electronic, Optical and Magnetic Properties},
  62(6):615--647, 1990.

\bibitem{Glazier}
James~A. Glazier, Steven~P. Gross, and Joel Stavans.
\newblock Dynamics of two-dimensional soap froths.
\newblock {\em Physical Review A}, 36(1):306--312, 1987.

\bibitem{Cohen-Addad2013_FlowinFoams}
Sylvie Cohen-Addad, Reinhard H{\"{o}}hler, and Olivier Pitois.
\newblock {Flow in Foams and Flowing Foams}.
\newblock {\em Annual Review of Fluid Mechanics}, 45(1):241--267, 2013.

\bibitem{Rosowski2020_NatPhys}
Kathryn~A. Rosowski, Tianqi Sai, Estefania Vidal-Henriquez, David Zwicker,
  Robert~W. Style, and Eric~R. Dufresne.
\newblock {Elastic ripening and inhibition of liquid–liquid phase
  separation}.
\newblock {\em Nature Physics}, 16(4):422--425, 2020.

\bibitem{Tanaka2000}
Hajime Tanaka.
\newblock {Viscoelastic phase separation}.
\newblock {\em Journal of Physics: Condensed Matter}, 12:R207--R264, 2000.

\bibitem{Tcholakova2011a}
Slavka Tcholakova, Ivan Lesov, Konstantin Golemanov, Nikolai~D. Denkov, Sonja
  Judat, Robert Engel, and Thomas Danner.
\newblock {Efficient emulsification of viscous oils at high drop volume
  fraction}.
\newblock {\em Langmuir}, 27(24):14783--14796, 2011.

\bibitem{Politova2018}
Nadya Politova, Slavka Tcholakova, Zhulieta Valkova, Konstantin Golemanov, and
  Nikolai~D. Denkov.
\newblock {Self-regulation of foam volume and bubble size during foaming via
  shear mixing}.
\newblock {\em Colloids and Surfaces A: Physicochemical and Engineering
  Aspects}, 539(December 2017):18--28, 2018.

\end{thebibliography}

\bibliographystyle{unsrt}

\noindent\textbf{Acknowledgements}\\
This study was partly supported by the Centre National d'\'Etudes Spatiale (CNES) through project "Hydrodynamics of wet foams". A.S. acknowledges V\'eronique Trappe and Eric Dufresne for interesting discussions. 

\noindent\textbf{Author contributions}\\
C.G., E.R. and A.S. designed the experimental study. C.G. carried out the experiments. C.G., J.M.I., E.R., A.P. and A.S. analysed and interpreted the experimental results. C.G., E.R. and A.S. wrote the manuscript with contributions from J.M.I and A.P. 

\noindent\textbf{Competing interests}\\
The authors declare no competing financial interests. 

\end{document}